\documentclass[12pt]{article}
\usepackage{psfig}
\usepackage{verbatim}
\usepackage{color}
\usepackage{rotate}
\usepackage{graphics}
\usepackage{epsf}
\usepackage{epsfig}
\usepackage{graphicx}
\usepackage{latexsym}
\usepackage{amscd}
\usepackage{amssymb}
\newcommand{\beq}{\begin{equation}}
\newcommand{\eeq}{\end{equation}}
\newcommand{\beqn}{\begin{eqnarray}}
\newcommand{\eeqn}{\end{eqnarray}}
\newcommand{\be}{\begin{equation}}
\newcommand{\ee}{\end{equation}}
\newcommand{\bea}{\begin{eqnarray}}
\newcommand{\eea}{\end{eqnarray}}
\newcommand\noi{\noindent}

\newcommand\eps\varepsilon

\def\lsim{\mathrel{\rlap{\lower4pt\hbox{\hskip1pt$\sim$}}
    \raise1pt\hbox{$<$}}}         
\def\gsim{\mathrel{\rlap{\lower4pt\hbox{\hskip1pt$\sim$}}
    \raise1pt\hbox{$>$}}}   
\def\Journal#1#2#3#4{{#1} {\bf #2}, #3 (#4)}

\def\PRL{\it Phys. Rev. Lett.}
\def\PRD{{\it Phys. Rev. D}}
\def\PRC{{\it Phys. Rev. C}}

\begin{document}

\hfill {LA-UR-02-1823}

\vspace*{2cm}
\begin{center}
{\Large
\bf 
Nuclear effects in the Drell-Yan process}
\end{center}
\vspace{.5cm}

\begin{center}
 {\large
J\"org Raufeisen\footnote{\tt email: jorgr@lanl.gov}}
\medskip

{\sl Los Alamos National Laboratory,
Los Alamos, New Mexico 87545, USA}
\end{center}

\vspace{.5cm}

\begin{abstract}
\noi
In the target rest frame and at high energies, Drell-Yan (DY) dilepton
production looks like bremsstrahlung of massive photons, rather than
parton annihilation. 
The projectile quark is decomposed into a series of Fock states. 
Configurations with fixed transverse separations in impact 
parameter space are interaction
eigenstates for proton-proton ($pp$) scattering. 
The DY cross section can then be expressed
in terms of the same color dipole cross section as DIS. 
We compare calculations in this dipole approach with E772 data
and with next-to-leading order
parton model calculations. 
This approach is
especially suitable to describe nuclear effects, since it allows one
to apply
Glauber multiple scattering theory.
We go beyond the Glauber eikonal approximation by taking into account
transitions between states, which would be eigenstates for a proton
target. 
We calculate nuclear shadowing at large Feynman-$x_F$ for DY in
proton-nucleus collisions and compare to E772 data. 
Nuclear effects on the transverse momentum
distribution are also investigated.
\end{abstract}

\clearpage

\section{DY dilepton production in $pp$ scattering}

\begin{figure}[th]
\centerline{
  {\scalebox{0.8}{\includegraphics{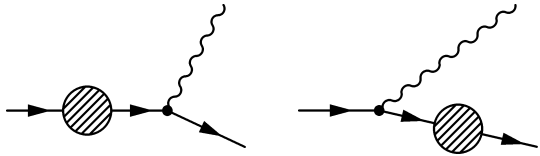}}}}
\caption{
  \label{bremsdy}
  {\em    A quark 
      (or an antiquark) inside the
      projectile hadron scatters off the target color field and radiates a
      massive photon. The subsequent decay of the $\gamma^*$ 
	into the lepton pair is not shown. }
}
\end{figure}

Although cross sections are Lorentz invariant, 
the partonic interpretation of the
microscopic process depends on the reference frame.  
As pointed out in \cite{boris},
in the target rest frame, Drell-Yan (DY)
dilepton production should be treated as bremsstrahlung, rather than parton
annihilation  
(see also \cite{bhq}). 
The space-time picture of the DY process in the target rest frame is
illustrated in Fig.~\ref{bremsdy}.  
A quark (or an antiquark) from the projectile
hadron radiates a virtual photon on impact on the target.

A salient feature of the rest frame picture of DY dilepton production is 
that at high
energies and in impact parameter 
space the DY cross section can be formulated in terms
of the same dipole cross section as low-$x_{Bj}$ DIS. 
The cross section for radiation of a virtual photon from a quark after
scattering on a proton, can be written in factorized light-cone form
\cite{boris,bhq,kst}, 
\be\label{dylctotal}
\frac{d\sigma(qp\to \gamma^*X)}{d\ln\alpha}
=\sum_{T,L}\int d^2\rho\, |\Psi^{T,L}_{\gamma^* q}(\alpha,\rho)|^2
    \sigma_{q\bar q}(\alpha\rho),
\ee
similar to the case of DIS.
Here, $\sigma_{q\bar q}$ is the cross section \cite{mpb} for scattering  a
$q\bar q$-dipole off a proton which depends on the $q\bar q$ separation 
$\alpha\rho$,
where $\rho$ is  the photon-quark transverse separation and $\alpha$ 
is the fraction of 
the light-cone momentum of the initial quark taken away by the photon.
For shortness, we do not explicitly 
write out the energy dependence of $\sigma_{q\bar q}$.
We use the standard notation for the kinematical variables,
$x_1-x_2=x_F$, $\tau=M^2/s=x_1x_2$, where $x_F$ is the
Feynman variable,
$s$ is the center of mass energy squared of the colliding protons and 
$M$ is the
dilepton mass. In (\ref{dylctotal}) $T$ stands for transverse and $L$
for longitudinal photons.

The physical interpretation of (\ref{dylctotal}) is similar to the DIS
case. The projectile quark is expanded in the
interaction eigenstates. We keep here only the first eigenstate,
\be
|q\rangle=\sqrt{Z_2}|q_{bare}\rangle+\Psi^{T,L}_{\gamma^* q}|q\gamma^*\rangle+\dots,
\ee
where $Z_2$ is the wavefunction renormalization constant for fermions.
In order to produce a new state the interaction must distinguish 
between the two Fock 
states, {\it i.e.} they have to interact differently. Since only the quarks
interact in both Fock components the difference arises from their relative displacement 
in the transverse plane.
 If $\rho$ is the transverse separation between the
quark and the photon, the $\gamma^*q$ fluctuation has a center of 
gravity in the
transverse plane which coincides with the impact parameter of the parent quark.
The transverse separation between the photon and the center of gravity is
$(1-\alpha)\rho$ and the distance between the quark and the center of 
gravity is
correspondingly $\alpha\rho$. Therefore, the argument of $\sigma_{q\bar q}$ is
$\alpha\rho$. More discussion can be found in \cite{last}.

The transverse momentum distribution of DY pairs
can also be expressed in terms of the dipole cross section \cite{kst}. 
The differential cross section is given by the 
Fourier integral
\bea\nonumber\label{dylcdiff}
\frac{d\sigma(qp\to \gamma^*X)}{d\ln\alpha d^2q_{T}}
&=&\frac{1}{(2\pi)^2}
\int d^2\rho_1d^2\rho_2\, \exp[{\rm i}\vec q_{T}\cdot(\vec\rho_1-\vec\rho_2)]
\Psi^*_{\gamma^* q}(\alpha,\vec\rho_1)\Psi_{\gamma^* q}(\alpha,\vec\rho_2)\\
&\times&
\frac{1}{2}
\left\{\sigma_{q\bar q}(\alpha\rho_1)
+\sigma_{q\bar q}(\alpha\rho_2)
-\sigma_{q\bar q}(\alpha(\vec\rho_1-\vec\rho_2))\right\}.
\eea
after integrating this expression over the transverse momentum
$q_{T}$ of the photon, one obviously recovers
(\ref{dylctotal}). 

\begin{figure}[th]
\centerline{
  {\scalebox{0.8}{\includegraphics{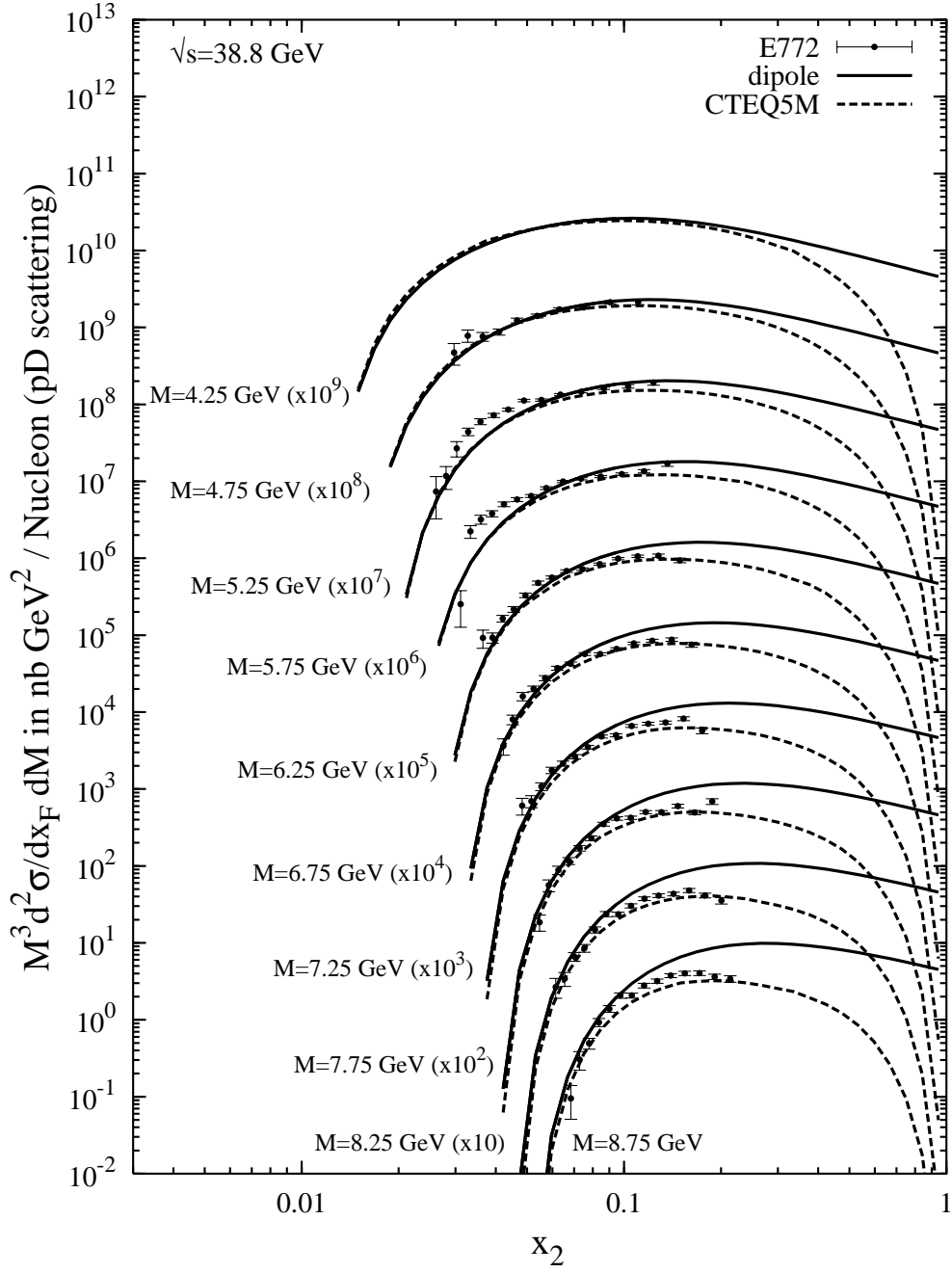}}}}
\caption{
  \label{dytotal}{\em
  The points represent the measured DY cross section in $p\,D$ scattering
  from E772 \cite{dydata}. 
  The solid curve is calculated in the dipole approach, while the 
NLO parton model calculation (using CTEQ5M parton distributions \cite{cteq5})
is shown as dashed curve. 
The dipole approach is valid only at small $x_2$.}
}
\end{figure}

The LC wavefunctions can be calculated in perturbation theory and are 
well known
\cite{bhq,last}.
The dipole cross section on the other hand is largely unknown.
Only at small distances $\rho$ it can be
expressed in terms of the gluon density. However, several 
successful parameterizations exist in the
literature, describing the entire function $\sigma_{q\bar q}(x,\rho)$, without
explicitly taking into account the QCD evolution of the gluon density.  
We use the
parameterization by Golec-Biernat and
W\"usthoff \cite{Wuesthoff1} for our calculations, Fig.~\ref{dytotal}.
This parameterization vanishes $\propto\rho^2$ at small distances, as
implied by color transparency \cite{mpb}
and levels off exponentially at large separations.
 
In Fig.~\ref{dytotal}, we compare to E772 data \cite{dydata} on
low $x_2$ DY dilepton production \cite{new}. 
Most of the data are quite well described 
without any $K$-factor, which does not appear in this approach since higher
order corrections are supposed to be parameterized in 
$\sigma_{q\bar q}(\rho)$. Moreover, the calculation in the dipole 
approach agrees with the next-to-leading order (NLO)
parton model calculation 
at low $x_2$. Note that the dipole approach is valid only at low $x_2$
\cite{last}.
At large $x_2$, this approach is not applicable and differs strongly
from the parton model calculation. 
The disagreement between the data and both of the calculations in some points
is probably due to systematic errors in the 
measured cross section.
Preliminary E866 data \cite{e866} agree well with the NLO parton 
model calculation.

\clearpage

\section{Proton-nucleus ($pA$) scattering}

The main advantage of the dipole approach is its easy generalization
to nuclear targets. Furthermore, it also includes some higher twist
effects that are important in multiple scattering, and it 
provides insight into the physical mechanisms underlying nuclear effects,
which are not easily accessible in the parton model \cite{pm}.

Shadowing in DY is an interference phenomenon due to multiple scattering of the
projectile quark inside the nucleus. In the target rest frame, where DY dilepton production
is bremsstrahlung of massive photons, shadowing is the Landau-Pomeranchuk-Migdal (LPM)
effect. 
These interferences occur (Fig.\ \ref{bremsdy}), because photons
radiated at different longitudinal coordinates $z_1$ and $z_2$ 
 are not
independent of each other. 
Thus, the amplitudes have to be added coherently. 
Destructive interferences can occur only if the longitudinal distance 
$z_2-z_1$
is smaller than the so called coherence length $l_c$, which is the time needed
to distinguish between a quark and a quark with a $\gamma^*$ nearby. 
It is given by the uncertainty relation,
\be\label{lc}
l_c=\frac{1}{\Delta
P^-}=\frac{1}{m_Nx_2}\,\frac{(1-\alpha)M^2}
{q_{T}^2+(1-\alpha)M^2+\alpha^2m_q^2}.
\ee
Here, $\Delta P^-$ is the light-cone energy denominator for the transition $q\to
q\gamma^*$ and $q_{T}$ is the relative transverse momentum of
the $\gamma^*q$ Fock state.
For $z_1-z_2>l_c$, the
radiations are independent of each other. 

An immediate consequence of this is that $l_c$ has to be
larger than the mean distance between two scattering centers in the nucleus
($\sim 2$ fm in the nuclear rest frame). Otherwise, the projectile quark could
not scatter twice within the coherence length and no shadowing would be observed.

We develop a Green function technique \cite{kst}, which allows one
to resum all multiple scattering terms,
similar to Glauber theory, and in addition treats the coherence length exactly.
The formalism is equivalent to the one proposed in \cite{Slava}
for the LPM effect in QED.
Our general expression for the nuclear DY cross section reads
\bea\nonumber
\frac{d\sigma(qA\to \gamma^*X)}{d\ln\alpha} & = &
A\,\frac{d\sigma(qp\to \gamma^*X)}{d\ln\alpha}- \frac{1}{2} {\rm Re}\int d^2b
\int_{-\infty}^{\infty} dz_1 \int_{z_1}^{\infty} dz_2
\int d^2\rho_1\int d^2\rho_2\,
\\ \nonumber
& \times &
\Bigl[\Psi_{\gamma^*q}\left(\alpha,
\rho_2\right)\Bigr]^*\,
\rho_A\left(b,z_2\right)\sigma_{q\bar{q}}\left(\alpha\rho_2\right)
{G\left(\vec \rho_2,z_2\,|\,\vec \rho_1,z_1\right)}\\
&\times &
\rho_A\left(b,z_1\right)\sigma_{q\bar{q}}\left(\alpha\rho_1\right)
\Psi_{{\gamma^*q}}\left(\alpha,\rho_1\right). 
\label{dyshadowing}
\eea

The first term is just $A$ times the single scattering cross section, where $A$ is the
nuclear mass number. The second term is the shadowing correction. The impact parameter is
$b$ and the nuclear density is $\rho_A$. The Green function $G$ describes, how the
bremsstrahlung-amplitude at $z_1$ interferes with the amplitude at $z_2$. 

To make the meaning of Eq.~(\ref{dyshadowing}) more clear, let us first
consider a limiting case for $G$. 
In the simplest case, the coherence length, Eq.\
\ref{lc}, is infinitely long 
and only the double scattering term is taken into account. Then 
$G\left(\vec \rho_2,z_2\,|\,\vec \rho_1,z_1\right)=\delta^{(2)}(\vec\rho_1-\vec\rho_2)$ and
one of the $\rho$ integrations can be performed. The $\delta$-function means that at very
high energy (infinite coherence length) the transverse size of the $\gamma^*q$ Fock-state
does not vary during propagation through the nucleus, it is frozen due to Lorentz time
dilatation. Furthermore, partonic configurations with fixed transverse separations in impact
parameter space were identified a long time ago 
\cite{mpb} in QCD as interaction eigenstates. 
This is the reason, why we work in coordinate space. Namely, in coordinate
space, all multiple scattering terms can be resummed and in the limit of 
infinite $l_c$ 
one obtains 
\be\label{frozen}
G^{frozen}
\left(\vec\rho_2,z_2\,|\,\vec\rho_1,z_1\right)=
\delta^{(2)}(\vec\rho_1-\vec\rho_2)
\exp\left(-\frac{\sigma_{q\bar q}(\alpha\rho_1)}{2}
\int_{z_1}^{z_2}dz\rho_A(b,z)\right).
\ee 
The frozen approximation is identical to eikonalization of 
the dipole cross section in Eq.\
(\ref{dylctotal}). Thus, the impact parameter representation allows a very simple
generalization from a proton to a nuclear target, provided the coherence length is
infinitely long.

At Fermilab fixed-target energies ($\sqrt{s}=38.8$ GeV for E772), 
this last condition is not fulfilled and one has to take a finite $l_c$ into account. The
problem is however, that $l_c$, Eq.~(\ref{lc}), depends on the relative transverse momentum
$q_{T}$ of the $\gamma^*q$-fluctuation which is the conjugate variable to the size $\rho$
of this Fock-state and therefore completely undefined in $\rho$-representation. The quantum
mechanically correct way to treat the $q_{T}^2$ in Eq.~(\ref{lc}) is to represent it by a
two-dimensional Laplacian $\Delta_{T}$ in $\rho$-space. The Green function which contains
the correct, finite coherence length and resums all multiple scattering terms fulfills a
two-dimensional Schr\"odinger equation with an imaginary potential,
\bea\nonumber
\left[{\rm i}\frac{\partial}{\partial z_2}
+\frac{\Delta_{T}\left(\rho_2\right)-\eta^2}
{2E_q\alpha\left(1-\alpha\right)}
+\frac{{\rm i}}{2}\rho_A\left(b,z_2\right)\,
\sigma_{q\bar{q}}\left(\alpha\rho_2\right)
\right]
{G\left(\vec \rho_2,z_2\,|\,\vec \rho_1,z_1\right)}\\ 
\qquad =
{\rm i}\delta\left(z_2-z_1\right)
\delta^{\left(2\right)}\left(\vec \rho_2-\vec \rho_1\right),
\label{sgl}
\eea
where $\eta^2=(1-\alpha)M^2+\alpha^2m_q^2$.
For details of the derivation, we refer to \cite{kst}.

The imaginary potential accounts for all higher order scattering terms. The Laplacian implies that
the Green function is no longer proportional to a $\delta$-function. This means the size
of the $\gamma^*q$ fluctuation is no longer constant during propagation through the nucleus.
One can say that an eigenstate of size $\rho_1$ evolves to an eigenstate of size
$\rho_2\neq\rho_1$, so transitions between eigenstates occur.

\begin{figure}[t]
  \centerline{\scalebox{0.6}{\includegraphics{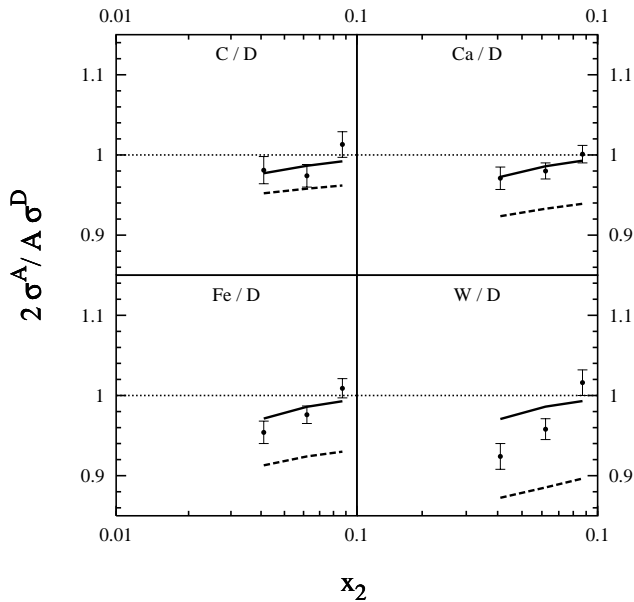}}}
    \caption{
      \label{e772}
      {\em Comparison between calculations in the Green function 
technique (solid curve)
and
      E772 data at 
center of mass energy {$\sqrt{s}=38.8$ GeV.} 
  for shadowing in DY. The dashed curve shows the eikonal 
(frozen) approximation, 
which is
not valid at this energy, any more.}
    }  
\end{figure}

Calculations 
with Eqs.~(\ref{dyshadowing}) and (\ref{sgl}) are compared to E772 data 
\cite{shadow} in Fig.~\ref{e772}. 
Note that the coherence length $l_c$ at E772 
energy becomes smaller than the nuclear radius. Shadowing vanishes as 
$x_2$ approaches $0.1$, because the coherence length becomes smaller 
than the mean internucleon separation.
It is therefore important 
to have a correct description of a finite $l_c$ in this energy range. 
The eikonal (frozen) approximation, Eq.~(\ref{frozen}), does not reproduce 
the vanishing shadowing toward $x_2\to 0.1$. 
The curves in  Fig.~\ref{e772} are somewhat different from the ones in
\cite{moriond}, because we used a different parameterization of the
dipole cross section.
Note that for heavy nuclei, energy loss \cite{eloss}
leads to an additional
suppression of the DY cross section.

\begin{figure}[t]
\centerline{
  \scalebox{0.5}{\includegraphics*{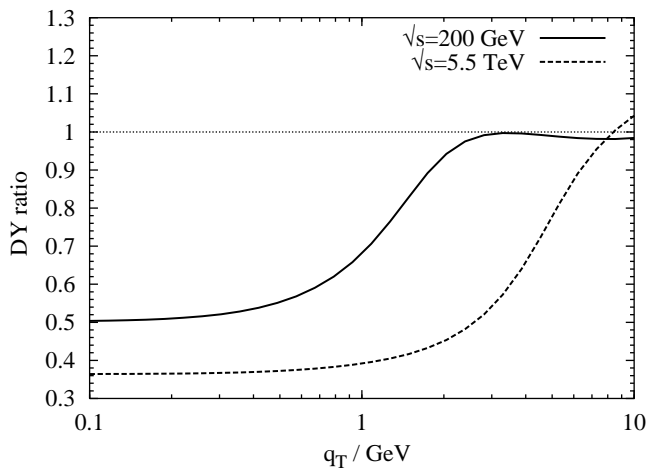}}
 }
\caption{\label{ratio}{\em
  Nuclear effects on the DY transverse momentum distribution at 
  RHIC and LHC for
  dilepton mass $M=4.5$ GeV and Feynman $x_F=0.5$. }
}
\end{figure}

Nuclear effects on the
$q_{T}$-differential cross section calculated at RHIC energy
are shown in Fig.\ \ref{ratio}.
See \cite{kst} for details of the calculation. 
The differential cross section is suppressed at small
transverse momentum $q_{T}$ of the dilepton, 
where large values of $\rho$ dominate. This suppression
vanishes at intermediate $q_{T}\sim 2$ GeV. 
The Cronin enhancement that one could expect in this intermediate
$q_T$ region \cite{moriond}
is suppressed due to gluon shadowing \cite{krtj}.  

A nuclear target 
provides a larger momentum transfer than a proton target and harder 
fluctuations
are freed, which leads to nuclear
broadening.
Note, that not the entire
suppression at low $q_{T}$ is due to shadowing. Some of the dileptons 
missing at low
$q_{T}$ reappear at intermediate transverse momentum. At very large 
transverse momentum nuclear
effects vanish.

\section{Summary}

We express the DY cross section in terms of the cross section for
scattering a $q\bar q$ dipole off a proton. This is the same dipole 
cross section that
appears in DIS. At low $x_2$ and for proton-proton scattering, 
calculations in the dipole approach agree
with calculations in the NLO parton model. Some E772 data points are not 
well described by either of the approaches, which is probably due to a 
systematic error in the measured cross section.

At very high energy, 
the dipole approach is easily extended to nuclear targets by eikonalization.
At lower fixed target energies (E772) the eikonal 
approximation is no longer valid, because
the size of a Fock state  varies during propagation through the nucleus. 
Therefore,
transitions between interaction eigenstates (i.e.\ 
partonic configurations with fixed transverse
separations)  
occur.

We develop a Green function technique,
which takes variations of the transverse size into account and 
resums all multiple scattering
terms as well. For light nuclei,
calculations with the Green function technique are in good 
agreement with DY
shadowing data from E772. For heavier nuclei, also energy loss becomes
important.

We have also calculated
nuclear effects in the transverse momentum distribution of DY pairs at RHIC
energy. The DY cross section is suppressed at low transverse momentum.
The expected Cronin enhancement at
intermediate $q_{T}\sim 2$ GeV is reduced because of gluon shadowing. 
Nuclear effects vanish at very large $q_{T}$.

\bigskip
{\bf Acknowledgments:}
The author wishes to thank the Gesellschaft f\"ur Schwerionenforschung (GSI),
Darmstadt, Germany for financial support during this summer school.
This work was supported in part by the U.S.\ Department of Energy 
at Los Alamos National Laboratory under Contract No.\ W-7405-ENG-38.

I am indebted to J\"org H\"ufner, Mikkel Johnson, Boris Kopeliovich,
Jen-Chieh Peng and Alexander Tarasov for valuable discussion.

\end{document}